\documentclass[12pt]{article}

\usepackage{sbc-template}
\usepackage{graphicx,url}
\usepackage[english]{babel}
\usepackage{listings}
\usepackage[utf8]{inputenc}  
\usepackage[T1]{fontenc}
\usepackage[table]{xcolor}
\usepackage{multirow}
\usepackage{float}
\usepackage[para]{footmisc}
\usepackage{parskip}
\usepackage{newfloat}
\usepackage{xspace}
\usepackage{url}

\newcommand{\angr}{\texttt{Angr} }
\newcommand{\tool}{\textsc{MalVerse}\xspace}

\DeclareFloatingEnvironment[fileext=frm,placement={!ht},name=Code]{listing}

\title{Malware MultiVerse: From Automatic Logic Bomb Identification to Automatic Patching and Tracing}

\author{
Marcus Botacin\inst{1},
André Grégio\inst{1}, 
}

\address{
Federal University of Paraná (UFPR) -
\email{\{mfbotacin, gregio\}@inf.ufpr.br}
}

\begin{document} 

\maketitle

\begin{abstract}
Malware and other suspicious software often
hide behaviors and components behind
logic bombs and context-sensitive execution
paths. Uncovering these is essential to react
against modern threats, but current solutions are
not ready to detect these paths in a completely
automated manner. To bridge this gap, we propose
 the Malware Multiverse (\tool), a solution able 
to inspect multiple execution paths via symbolic
execution aiming to discover function inputs and 
returns that trigger malicious behaviors. \tool
automatically patches the context-sensitive 
functions with the identified symbolic values
to allow the software execution in
a traditional sandbox. We implemented \tool 
on top of \angr and evaluated it
with a set of Linux and Windows evasive
samples. We found that \tool was able to generate 
automatic patches for the most common evasion 
techniques (e.g., \texttt{ptrace} checks). 

\end{abstract}

\section{Introduction}
\label{sec:intro}

Reverse engineering is one of the most common tasks of malware's
binary analysis. Security experts invest a lot of time trying to
infer the goals of a malicious binary and how to bypass
anti-analysis code. Logic bombs---malware reliance on specific
conditions to be triggered---are one of the most challenging
techniques to counter (e.g., context-sensitive malware run only
under a given machine, timezone, and with/without a
debugger~\cite{bomb3}), and have been often seen in the
wild~\cite{bomb1,bomb2}. However, there is still a lack of automated
solution to defuse them.

A promising (and often used) technique to identify logic bombs is symbolic analysis,
since it allows the execution of multiple paths to eventually discover those that
trigger malicious behaviors. Symbolic executors work by replacing concrete execution 
values with symbolic values that can be constrained according to the followed path. 
Several research works propose to assist malware analysis with symbolic 
execution~\cite{10.1007/978-3-319-75160-3_22, Alsaleh2019,
Brumley2008}, but none provides a completely automated solution for
this task. Therefore, significant
challenges have to be overcome before the successful development of an automated symbolic 
execution-based tool for logic bombs handling, such as: (i) automatically
identification of entry and exit points of the analyzed application (most previous work
rely on human analysts to it); (ii) tracking information among multiple 
processes (most symbolic executors do not support forking or Inter Process
Communication--IPC~\cite{8443109}, both commonly used by malware);
(iii) time issues, as even if the tools supported forking/IPC,
symbolic analysis of whole binaries tend to be extremely long (hours);
and (iv) intent inference, as once multiple paths are discovered,
there is no automated way to claim which one is particularly
malicious.
 
Aiming to develop an automated solution for the analysis of malware
samples armored with logic bombs, we propose \tool (the Malware
Multiverse). The main goal of \tool is to allow efficient symbolic
execution on malware samples and traverse their multiple paths (or
``universes'') to identify logic bombs, and then
generate patches to these malware to allow for binary inspection on typical 
sandbox solutions. With \tool, we intend to provide a fully automated
solution that mitigates the cost of symbolic analysis, whereas
allowing the inspection of complex, modular binaries.

\tool is developed on top of \angr~\cite{shoshitaishvili2016state},
relying on its premise that the whole control flow is defined by
function returns, either internal (binary) or external (library)
calls. Upon that premise, \tool speeds up analyses by hooking all
function calls and replacing them with newer versions
that returns a symbolic variable instead of actually executing the native 
function (both for libraries and for internal binary's calls): it makes 
analyses much faster \textbf{and} considers the actual decision nodes.
After enumerating all paths, \tool applies a Bayesian decision model to each function
in a given path to filter the potentially malicious ones. This model is trained with multiple traces
of malware samples and allows the identification of malicious-related functions
(e.g., what is the probability that the binary function \textit{X} is 
associated to malicious activities given that it originally was programmed to invoke
\texttt{socket} and \texttt{send}?). After selection, \tool restarts
the analysis procedures only for the filtered functions until all
symbolic values are identified.

After the retrieval of all symbolic values, \tool produces patches for functions 
invoked on nodes that lead to the malicious paths (e.g., making that
\texttt{IsDebuggerPresent} always returns \texttt{False} to force the
execution through a non-evasive path).
The patching mechanism is implemented on top of \angr's decompiler, but instead
of decompiling the original function, it replaces the function body with new
statements to reflect the identified symbolic values. The decompiler generates
fully functional code that might be directly compiled into a library to be
injected in the original process (e.g., via \texttt{LD\_PRELOAD} or \texttt{DLL
Injection}). This way, \tool reaches malicious states identified in the
symbolic execution while running the binary in traditional sandbox solutions,
consequently overcoming existing limitations of symbolic executors, such as handling
forking and IPC.

Our experiments show that \tool is able to produce patches
that overcome many popular logic bombs and context checks, 
such as skipping debuggers (e.g., via \texttt{ptrace}
and \texttt{IsDebuggerPresent}) and stalling code (e.g.,
bypassing \texttt{sleep} functions). 

In summary, our contributions are threefold:
\textbf{i)} We revisit the problem of logic bombs identification and pinpoint the challenges to the automation and the escalation of malware analysis using symbolic executors. We expect that their identification and discussion might guide future developments to overcome them;
\textbf{ii)} We introduce \tool, a solution that automatically identifies malicious execution paths through symbolic execution and produces patched libraries to allow the binary inspection
in standard tracing solutions (e.g., sandboxes);
\textbf{iii)} We evaluate \tool with \texttt{Linux} and \texttt{Windows} evasive malware, showing its effectiveness in discovering multiple infection paths.

The remainder of this paper is organized as follows:
in Section~\ref{sec:bombs}, we revisit the logic bombs
identification problem in the context of malware evasion
attempts, and delve into the challenges faced to automatically 
overcome those threats;
in Section~\ref{sec:impl}, we introduce \tool's design
and implementation;
In Section~\ref{sec:eval}, we evaluate \tool with real
malware samples;
In Section~\ref{sec:disc}, we discuss the impact of
our findings and the \tool's limitations;
In Section~\ref{sec:rel}, we present the related work so as to
better position our contributions to the literature; finally, we draw our conclusions in Section~\ref{sec:conc}.

\section{Logic Bombs \& Symbolic Execution}
\label{sec:bombs}

During an infection, modern malware samples do not immediately
proceed to their malicious steps. Instead, they probe the
environment to verify whether it is safe to run or not (e.g., if they are running
inside an analysis environment, such as a virtual machine, emulator, debugger). If so, they refuse to execute or exhibit a non-malicious 
behavior, thus not revealing their real intents to analysts. 
This way, malware go on performing their actions on
victims machines without being noticed by the security
companies.
Modern malware also fingerprint the environment before 
executing malicious actions. This allows them to have unique 
identifiers to account the number of successful infections, 
identify which payloads will be used to exploit specific software
versions, and even attack only targeted users and/or machines
(e.g., by checking the system language, the local IP
addresses, the system architecture, so on).
All verification routines like those, which trigger a malicious
execution only in specific cases, are known as logic bombs, and their identification is essential to allow analysts develop countermeasures against armored malware.

Identifying these logic bombs may not be trivial. The
more time an analyst spends reverse engineering this type of sample,
the greater the attack opportunity window for malware affecting users in-the-wild. Therefore, the development of tools to automatically defuse these logic bombs is essential
to reduce security solution's response time.
The first attempts to automatically defuse logic bombs
were through fuzzing approaches, which generate multiple
distinct inputs to force malware execution via multiple
paths~\cite{8576654}, then trace further execution.
The major drawback of these solutions is that they depend
on ``luck'' to find the malicious paths, and very specifically
targeted malware might still remain hidden in hard-to-find
paths. In addition, these approaches are not fully able to
provide guarantees about how many paths can lead to a
malicious state.
If an analysis stops right after the first
path is found, it might skip important secondary 
paths to a target. Those secondary paths can often
be found via additional search time. In some
cases, those skipped paths might not be armored
with any anti-analysis technique, thus they would allow 
an analyst to circumvent the anti-analysis techniques present in the main/first path by forcing the flow via
those secondary paths.

The most promising current approach for logic bomb detection
is the use of symbolic executors. They lift a binary to an
intermediate representation so as to mathematically model 
the possible execution paths. This kind of model makes the execution
aware of the number of possible paths and allows one to
clearly explain how a input that reached the found path 
was generated. Unlike fuzzers, symbolic executors do
not operate on concrete values, but on symbolic values
that will assume a concrete value only in the end of the 
execution. According to the executed instructions over
a path, the symbolic executor adds constraints to the possible
values that these symbolic variables can assume. The set
of constraints that a symbolic executor is operating on a
given path is called state. Modern symbolic executors can
execute multiple states in parallel by forking states
each time a branching condition is found.

As an example for the case of logic bombs, consider the
symbolic execution of a context-sensitive malware from its
initial state. When a context-sensitive function is invoked 
(e.g., \texttt{IsFeaturePresent}), the symbolic executor 
supplies a symbolic value as return of this function. This 
symbolic variable has no concrete value or constraints at
this point. As the execution proceeds, this value will be
further referenced in a comparison (e.g., \texttt{if processor
feature is present, then malicious; else evade}). When the
symbolic executor finds this branching condition, it creates
two new states. The first state has the symbolic variable
with the condition \texttt{feature is present}. The second
state has the symbolic variable with the condition \texttt{feature
not present}. The two states are then independently analyzed.
If new branching conditions are found, new constraints are added
(to the same or new symbolic variables). 

Modern symbolic executors require analysts to specify a
target execution point. In this sense, three
scenarios might take place for the above example: (i) both
states reached the same breakpoint, which means that the value
returned by that function did not effectively affect the
execution control flow; (ii) one of the states reached the
breakpoint, which means that the control flow was dependent
of the return value assigned by that function; and (iii)
none of the states reached the breakpoint, which means that
the specified execution path was unfeasible. This might
happen, for instance, when contradictory constraints
are imposed (e.g., \textit{x} should be  greater than 0,
and \texttt{x} should be smaller or equal to zero).

Logic bombs can be identified by repeating the aforementioned procedure for all
function calls that an analyst identifies as prone to be used for
context identification. This was proposed by many previous
work~\cite{10.1007/978-3-319-75160-3_22,Alsaleh2019,Brumley2008}, with the drawback of all of them being semi-automated solutions~\cite{10.1145/3098954.3105821,9018111}. With \tool, we aim to develop a fully automated solution to identify logic bombs, whereas addressing some of the significant challenges (C) of performing symbolic execution on a binary:

\noindent \textbf{C1. Identifying Entry and Exit Points:} Current symbolic
executors require the analyst to specify the execution entry
point (initial state) and exit points (execution target). Specifying
an improper entry or exit point might lead to unsuccessful analysis
results, such as unreachable paths and the hidden execution
paths. Current binaries present multiple functions and 
possibilities and previously proposed solutions relied on the analyst's knowledge for this task.

\noindent \textbf{C2. Overcome Unsupported Methods:} although fuzzing
solution actually execute the function calls, symbolic executors
have to model the interactions performed by the calls. This results
in the fact that many functionalities are often unsupported by
these solutions due to the required modeling complexity. For instance, interactions such as creating a new process and
Inter-Process Communication (IPC) are often not supported by
many symbolic executors~\cite{8443109}. Unfortunately, these are techniques
very common in malware samples, thus a malware analysis solutions
need to overcome this limitation.

\noindent \textbf{C3. Performance Bottlenecks:} even if all malware
required functionalities are supported by the symbolic executor,
analysis are challenging due to their expensive computational
cost, both in terms of processing time as well as in resource
requirements. Symbolic executors have to analyze an exponential
number of states, often in parallel, storing deep constructions
in memory, and interpreting functions without actually executing
them. This makes analysis to take a significant time. It is not
rare to observe cases in which analysis takes multiple minutes~\cite{10.1007/978-3-319-75160-3_22} or even hours. 
This is a major limitation for solutions aiming to reduce 
the security solution's response time. The solution would
not be competitive if it takes  the same time than an analyst
to perform the task.

\noindent \textbf{C4. Identifying Suspicious Paths:} after the analysis
is finished and the symbolic executor provided all possible execution
paths to reach a target, there is also the remaining challenge of identifying what paths are malicious, since reaching the same final
state does not mean that the actions performed in the path had all
the same side-effects in the system. This task is often guided
by the analyst's knowledge and no previous solution automatically
classified found states as malicious or benign.

\section{Design and Implementation of \tool}
\label{sec:impl}

\noindent \textbf{Threat Model.} \tool's goal is to automate the
main tasks performed by analysts to bypass evasive malware checks.
We are aware that there are multiple ways to deceive analysis, making some techniques not bypassable at all. However, saving analysts time already helps in incident response, and to accomplish that, we automated some of the most common classes of
evasion techniques. In this sense, we believe that a divide-and-conquer strategy
is the best approach to follow in the long-term to solve the complex
automatic patching problem. In this work, we tackle
the case in which we are able to control function returns,
both internal (binaries) and external (libraries) ones. Therefore,
\tool assumes that all control flow decisions are due to function
return values. \tool was developed on top of \angr, due to the latter being a very extensible solution, but it can be implemented on top of 
any symbolic execution solution. Similarly, for the sake of 
simplicity, we exemplify our patching mechanism using standard 
tracing solutions (e.g., \texttt{strace}), but it can be applied 
to any debugger or sandbox.

\noindent \textbf{Design.} Previously, we presented the challenges
involved in the development of a fully automated solution to defuse
logic bombs within malware samples. 
Our key insight to overcome those challenges is that it is not
necessary to handle the whole complexity of binary analysis
to provide a first look on malware behavior. Therefore, we introduce a
set of heuristics, methods, and strategies that allows automated
triage of logic bomb-armored malware samples, inspired by the
observation present in~\cite{10.1145/3375894.3375895}, i.e., that
malware decompilation could be eased if only the actually executed
code during an analyst's
debugging session were considered. In this work, we extend this idea by decompiling the symbolic values that make a function to actually reach targeted code portions, and executing them on a typical sandbox solution to retrieve more precise execution
traces. We overcome logic bombs defusing challenges with the following design (D) choices:

\noindent \textbf{D1. Identifying Entry and Exit Points.} We leveraged
\angr's analysis capabilities to generate a Call Graph (CG) of
the inputted binary. We identify in this CG the first function
of the binary that invokes other functions as the main function.
The address of this function is set as the analysis entry point.
Similarly, we identify the return sites of this function as target
addresses. If any of the return addresses is reached, we consider 
that the function was executed.

\noindent \textbf{D2. Automatic Function Modeling.} Library's function
calls were modelled via \angr's \texttt{SimProcedures}, which were all set to return symbolic values. Manually coding/adapting \angr's SimProcedures for all APIs is
laborious, thus we developed an automated procedure
to help in this task.
We developed a Web crawler that identifies function 
prototypes through the Internet and automatically generates 
\texttt{SimProcedures} for them (discussed below).

\noindent \textbf{D3. Keeping Invocation History.} We need to keep
track of function invocations during a given path to be able
to compare two paths and identify in each point the execution
diverted. To do so, we added a \texttt{SimInvocationHistory}
to the \angr state. It complements already-existing events
and action histories with invocation information. Therefore,
each time a procedure is invoked, it adds itself (and its
symbolic values) to the history. In the end of the execution, 
we can compare whether two invocations \textit{concretize} to 
the same values or not.

\noindent \textbf{D4. Overcoming Performance Bottlenecks.} We rely on
the hypothesis that the whole control flow is due to function
returns to speed up analysis procedures. Once we identify a
main function, we replace all functions invoked by
them (according to the CG) by a procedure that only returns 
a symbolic value, that can assume any value originally returned
by this function. This does not break the execution flow if 
our premise is valid and speeds up analysis by avoiding the 
analyzer to effectively dig into that function. If one of 
these functions is identified as the root cause of a control 
flow diversion, the analysis restarts only for that function,
recursively performing this strategy until the actual root 
cause is identified.

\noindent \textbf{D5. Identifying Suspicious Paths.} The symbolic
executor often finds multiple different paths to reach the
return condition. In this case, we need to identify which
ones are suspicious and/or malicious. Once we identify them,
we force our analysis to proceed via them, thus ensuring that
we are inspecting malicious cases and not error conditions. We identify whether our execution flow should go through a given function present on the CG by applying a Bayesian decision procedure. 
We trained a model with multiple malware traces,
as presented in a previous study~\cite{sbseg2019},
and applied this model to each function on the CG. We query if a given function
should be considered on not (e.g., Should a malicious path
goes through \texttt{X} given that \texttt{X} invokes
\texttt{fork+exec}?). We discard the paths that seem to
not exhibit malicious behaviors. Once two similar paths
that traverse the sames functions are identified, they 
are aligned via the \textit{concretized} values for each 
state. The first unmatched and/or mismatched function is 
considered as the diversion root cause. 

\noindent \textbf{D6. Overcoming Unimplemented Features.} We are
aware that some constructions, such as IPC, are hard to model
in symbolic executors. Therefore, we did not try to do that.
Instead, we focused on discovering whether the path that
contains an IPC should be traversed or not, and what are the conditions that trigger that execution. After this path is
identified, we opted to run it on a standard sandbox,
with full support to IPC. Therefore, each time a diversion
point is identified, we produced a patch for that function. This patch may then be compiled into a library and injected in the binary while running in a sandbox, which forces the execution through the identified path and allows the sandbox to collect full binary information about the complex behaviors.

\noindent \textbf{D7. Code Decompilation.} We developed our patching
system on top of the \angr decompiler. However, since it was
originally designed to decompile actual code (and our goal is to
produce a patch based on the found symbolic values), we have
to adapt it to generate code in distinct manners: when a patch
is requested to the decompiler, it replaces the original function variables with the \textit{concretized} values from symbolic variables; it also replaces the function body with new instructions that reference the new variables. Since the compiler preserves the original function prototypes and arguments, the output is a working code that can be compiled into a library.

\noindent \textbf{Implementation.} \tool required a lot of engineering work on \angr's internal code to support the newly added capabilities (\tool's code is available to anyone interested in checking these modifications). Below, we describe the changes performed to directly support \tool's approach.
A key part of \tool operation is its interaction with
\angr \texttt{Simprocedures}. We developed an automated
solution to generate \texttt{Simprocedures}: it crawls Internet websites for function prototypes, parses them, creates symbolic variables to be returned, and produces fully functional code.

Code~\ref{code:introspect} shows the generation solution
in action. It takes the \texttt{ptrace} function
as argument and generates a procedure for it. Notice that
the function protype was parsed so as to add the function
arguments to the procedure. The returned variable \textit{rval} 
is symbolic, thus allowing \angr to decide whether it should 
return or false according to the context. Line 4 shows that
the procedure adds itself to the history of invoked procedures.
It allows \tool querying this list at any time to check what
were the parameters provided to and returned by this function 
at any time. 

\begin{center}
\begin{minipage}{.95\textwidth}
\begin{lstlisting}[frame=single,label=code:introspect, caption={\textbf{API Crawler.} Simprocedure is automatically generated.}]
python3 simprocgen.py -t Linux -a ptrace --symbolize 
class ptrace(angr.SimProcedure):
	def run(self, request, pid, addr, data):
	    self.state.history.add_simproc_event(self)
		return self.state.solver.BVS(ptrace, 64, key=('api', ptrace))
\end{lstlisting}
\end{minipage}
\end{center}

\noindent \textbf{Bayesian Model.} We trained a Bayesian
classifier with traces obtained from the execution
of 5,000 benign samples (from fresh OS installations
of Windows 8/System32 files and Ubuntu 18/bin files)
and malware samples. The goal of this classifier is to 
identify the probability that a given function import is 
related to malicious behavior. 
For instance, the import of the \texttt{socket} function
is largely more prevalent (more than 70\%) in malware 
samples than in benign samples. Therefore, a binary 
function that references the \texttt{socket} function 
will be flagged as required to be traversed. We considered 
a threshold of 70\% for the Bayesian classifier confidence,
having the \texttt{socket} API as reference, as network
communication is a key step for a malware infection
process. If a binary function references more than 
one library function, we use the highest calculated probability 
to consider whether the binary function is required to be
traversed or not.

\section{Evaluation}
\label{sec:eval}

\noindent \textbf{\tool operation through examples.} A popular
return-based method used by malware to evade analysis 
is to check if a debugger is attached to the running process
and, if true, abort the execution. In the \texttt{Linux} environment,
it can be done by trying to attach the debugger to itself
via the \texttt{ptrace} call, which is denied if a debugger
was previously attached~\cite{trick1}, as shown in
Code~\ref{code:ptrace1}.

\begin{center}
\begin{minipage}[b]{.49\textwidth}
\begin{lstlisting}[frame=single,label=code:ptrace1, caption={\textbf{Debugger Evasion.} Analysis evadade if debugged.}]
if(ptrace(PTRACE_TRACEME)==-1){
    evade();
\end{lstlisting}
\end{minipage}
\begin{minipage}[b]{.49\textwidth}
\begin{lstlisting}[frame=single,label=code:ptrace2, caption={\textbf{Patched Ptrace.} The
debugger check will always fail.}]
long ptrace(int request, ...){
    return 0x0;
\end{lstlisting}
\end{minipage}
\end{center}

If the \texttt{ptrace Simprocedure} is instrumented to return a symbolic
value, as shown in Code~\ref{code:introspect}, it will be constrained by
\angr to a value different of \texttt{-1} (debugger present). Thus, \tool
can them decompile a patched version of the function returning the
identified value, as shown in Code~\ref{code:ptrace2}.

While patching a single function might be effective in specific cases, evasive malware might employ nested techniques in practice. In those 
cases, \tool identifies the distinct functions to be patched and
decompiles them together, as shown in Code~\ref{code:debugme} (regarding to the bypass of the \texttt{DEBUGME} sample~\cite{trick2}).

\begin{center}
\begin{minipage}{.9\textwidth}
\begin{lstlisting}[frame=single,label=code:debugme, caption={\textbf{Nested
Patched Functions.} Multiple functions are decompiled together.}]
long ptrace(int request, pid_t pid, void *addr, void *data){
    return 0x0;
int memcmp(const void *s1, const void *s2, size_t n){
    return 0x0;
\end{lstlisting}
\end{minipage}
\end{center}

The same debug evasion techniques can also be implemented in \texttt{Windows}.
Samples may query the \texttt{IsDebuggerPresent} API directly to verify if
they are being debugged, as shown in Code~\ref{code:win1}.

\begin{center}
\begin{minipage}{.8\textwidth}
\begin{lstlisting}[frame=single,label=code:win1, caption={\textbf{Checks on Windows.} 
Sample evades debugger or checks for an specific processor feature.}]
if(IsDebuggerPresent()==TRUE || 
   IsProcessorFeaturePresent(RANDOM_FEATURE)==FALSE){
    evade();
\end{lstlisting}
\end{minipage}
\end{center}

\tool can identify this debugger check and produce a patch that allows
its bypass, as shown in Code~\ref{code:win2}.

\begin{center}
\begin{minipage}[b]{.38\textwidth}
\begin{lstlisting}[frame=single,label=code:win2, caption={ 
Debugger is never present.}]
BOOL IsDebuggerPresent(){
    return 0x0;
\end{lstlisting}
\end{minipage}
\begin{minipage}[b]{.59\textwidth}
\begin{lstlisting}[frame=single,label=code:win3, caption={\textbf{Secondary Path.} 
The processor feature should always be present.}]
long IsProcessorFeaturePresent(long v){
    return 0x1;
\end{lstlisting}
\end{minipage}
\end{center}

The sample exemplified in Code~\ref{code:win1} presents a secondary
path that leads to evasion: it happens when a debugger is not present, but a given
processor feature is missing. In this case, \tool also identifies the
secondary path and decompiles the patched version, as shown in
Code~\ref{code:win3}.

In addition to the distinct paths that can be traversed independently,
some samples present evasion paths that depends on the nested invocation
of the same function. Code~\ref{code:ptrace3} illustrates the
\texttt{double ptrace} technique~\cite{trick3}, in which the
\texttt{ptrace} function is invoked twice to ensure that the 
bypass previously presented does not succeed. In this case,
during a run outside a debbuger, the first \texttt{ptrace} will
succeed on attaching the sample, but the second should fail.
If the sample runs under a patched version that always return zero,
the check for the second call will fail, incurring into an evasion. \\

\begin{center}
\begin{minipage}{.9\textwidth}
\begin{lstlisting}[frame=single,label=code:ptrace3, caption={\textbf{Double Ptrace.} 
This check relies on internal function states and side-effects.}]
if(ptrace(PTRACE_TRACEME)==0 && ptrace(PTRACE_TRACEME)==-1){
    evade();
\end{lstlisting}
\end{minipage}
\end{center}

\tool must keep track of the invocation order to bypass this type of check. To do so, it creates a global variable to count the number of invocation and associates a distinct return value to each invocation, as
shown in Code~\ref{code:ptrace4}. For the first invocation, \tool returns
that the function call succeeded (returning 0), whereas for the nexto one, it
returns that the function call failed (return value of -1). We highlight that
all the control flow code (counter variables and IFs) are automatically
generated by a version of the \angr decompiler instrumented with \tool code.

\begin{center}
\begin{minipage}{.9\textwidth}
\begin{lstlisting}[frame=single,label=code:ptrace4, caption={\textbf{Stateful Patches.} Each function return is associated to a distinct invocation.}]
#include<sys/types.h>
static int angr_global_var = 0;
long ptrace(int request, pid_t pid, void *addr, void *data){
    angr_global_var = angr_global_var + 1;
    if (angr_global_var == 1){
        return 0;
    }if (angr_global_var == 2){
        return -1;
\end{lstlisting}
\end{minipage}
\end{center}

This same strategy can be applied to bypass samples that make use of stalling
code to cause sandbox timeouts. Code~\ref{code:clock1} illustrates a sample
that checks CPU ticks to ensure the call of the sleep function was 
not replaced by a fake one that does not stall the execution.

\begin{center}
\begin{minipage}[b]{.49\textwidth}
\begin{lstlisting}[frame=single,label=code:clock1, caption={\textbf{Stalling Code.} 
The code checks if the process really spent cycles sleeping.}]
int main()
{
	clock_t t0 = clock();
	sleep(SLEEP_TIME);
	clock_t t1 = clock();
	if((t1-t0)>SLEEP_CLOCKS){
		malware();
	}else{
		goodware();
    }
}
\end{lstlisting}
\end{minipage}
\begin{minipage}[b]{.49\textwidth}
\begin{lstlisting}[frame=single,label=code:clock2, caption={\textbf{Patched Stalling
Code.} Both the sleep and clock functions reflect the imposed constraints.}]
unsigned int sleep(unsigned int seconds){
	return 0;
int angr_global_var = 0;
clock_t clock(void){
    angr_global_var = angr_global_var + 1;
    if (angr_global_var == 1){
        return 0x0;
    if (angr_global_var == 2){
        return 0xb;
\end{lstlisting}
\end{minipage}
\end{center}

To bypass this type of technique, \tool patches the \texttt{sleep}
function to immediately return, and the clock functions to reflect
the constraints expected by the malware sample, as shown in Code~\ref{code:clock2}.
This allows the inspection of the sample without waiting for the stalling code
execution time.

Finally, in some cases, values returned by functions only
indirectly control the execution flow. Code~\ref{code:cwd1}
exemplifies a context-sensitive malware that is only activated
when executed from a given path. Although it relies on the value
returned by \textit{getcwd}, it is stored on a memory 
position indeed returned as a pointer by this function.

\begin{center}
\begin{minipage}{.65\textwidth}
\begin{lstlisting}[frame=single,label=code:cwd1, caption={\textbf{Context-Sensitive Malware.}
It is only malicious when executed from a given path.}]
int main(){
    if(strcmp(getcwd(NULL,0),"BOMB")==0){
		malware();
	}else{
		goodware();
\end{lstlisting}
\end{minipage}
\end{center}

This type of stateful function requires implementing internal logic in the
\texttt{Simprocedure}, as shown in Code~\ref{code:cwd2}. The function should
allocate a memory buffer to return a valid pointer, whereas the allocated memory should
host a symbolic variable that stores the actual path.

\begin{center}
\begin{minipage}{\textwidth}
\begin{lstlisting}[frame=single,label=code:cwd2, caption={\textbf{Simprocedure Code.} 
The procedure allocates memory and stores a symbolic value there.}]
class getcwd(angr.SimProcedure):
  def run(self, buf, size):
    self.state.history.add_simproc_event(self)
    val = self.state.solver.BVS('getcwd', 64, key=('api', 'getcwd'))
    malloc = angr.SIM_PROCEDURES['libc']['malloc']
    addr = self.inline_call(malloc, 100).ret_expr
    self.state.memory.store(addr, val)
    return addr
\end{lstlisting}
\end{minipage}
\end{center}

In this case, it is ineffective to naively leverage \tool to decompile the \texttt{getcwd} function with the obtained concrete values, since the return value would point
to an invalid memory region if run out of the scope of the symbolic executor (as
shown in Code~\ref{code:cwd3}).

\begin{center}
\begin{minipage}{.75\textwidth}
\begin{lstlisting}[frame=single,label=code:cwd3, caption={\textbf{Naive Decompilation.} 
The code points to an invalid memory value.}]
int getcwd(char* var0, unsigned long var1){
    return 0xc0000f20; //needs to point to "BOMB"
\end{lstlisting}
\end{minipage}
\end{center}

To allow the decompilation of a fully functional piece of software,
\tool produces a patch for the targeted function and for the \texttt{main} function, in order to preload it with code that allocates
valid memory (as shown in Code~\ref{code:cwd4}).

\begin{center}
\begin{minipage}{.6\textwidth}
\begin{lstlisting}[frame=single,label=code:cwd4, caption={\textbf{Smart Patching.} 
In addition to patching the function, the main function is also preloaded with code
to allocate valid memory.}]
#define STR "BOMB"
void *addr;
static void init (void){
  addr = (char *) malloc (100);
  strcpy (addr, STR);
char * getcwd (char *buf, size_t size){
  return addr;
\end{lstlisting}
\end{minipage}
\end{center}

\section{Discussion}
\label{sec:disc}

\noindent \textbf{Hidden paths in distinct 
Operating Systems.} Since \tool was developed 
on top of \angr, it can be applied to binaries 
targeting multiple OSes, as presented in
Section~\ref{sec:eval}. Despite this fact, we 
noticed that \tool application to some OSes 
might be more effective than others, due to their very nature. In particular, applying \tool to Windows might be more effective than to 
Unix-based OSes: Windows presents a myriad of 
APIs to query context-specific information 
(e.g. \texttt{IsDebuggerPresent()}, whereas
Unix-based systems often perform context 
acquisition by directly querying the
filesystem (e.g., \texttt{/proc}), consequently requiring an alternate approach (e.g., a model of the accessed file) than \tool's  to be applied in these cases.

\noindent \textbf{The impact of API models.} \tool relies
on \angr's \texttt{Simprocedures} to model API behavior, and
the better you model these behaviors, the obtained results will be less inaccurate.
Currently, \tool instruments \texttt{Simprocedures} to return symbolic values, but we plan to also instrument function argument to increase the analysis capabilities (left as future work).

\noindent \textbf{Premises and Heuristics.} \tool's major
assumption is that the whole control flow is due to function
returns. Although we showed in this paper that it allows the bypass of many evasion techniques, we are aware that
internal function states can also affect the control flow.
Therefore, we intend to consider these cases in a future version, aiming at a more complete treatment of evasion cases. Similarly,
although our heuristic approach has shown to be effective on
the identification of malicious paths, we are aware that they can be
evaded by attackers (e.g., with the addition of suspicious functions
to all decision nodes, causing a path explosion). To address that, we will start to investigate the robustness of \tool heuristic procedures.

\section{Related Work}
\label{sec:rel}

\noindent \textbf{Discovering malware
secrets.} Multiple authors proposed
distinct approaches that leverage
symbolic execution and control flow
analyses to discover hidden
paths in malware execution, such as 
forcing the malware to take a given
path~\cite{8576654}. This might be useful
not only to discover intentionally-covered
paths but also to reconstruct Control 
Flow Graphs (CFG)~\cite{10.1145/3368926.3369702}
when a binary's control flow is 
obfuscated~\cite{10.1145/2810103.2813663}.
A major problem is that the symbolic 
executors leveraged for such executions
are vulnerable to anti-analysis 
tricks~\cite{10.1145/3371307.3371309}
that might hinder the analysis procedures.
To mitigate this problem, symbolic execution
is often performed along other techniques,
such as fuzzing~\cite{driller}. In this work,
we also present a dual-step approach, in which
symbolic execution is first employed to discover
control flow conditions and further analyses
are scaled via the application of traditional
sandbox-based execution. Scaling
analyses is important because it allows the
inspection of real malware samples. The closest
work to ours shows that it is possible to
recover even C\&C's control parameter
from the symbolic execution of a bot 
sample~\cite{10.1007/978-3-319-60080-2_12}. 
The major limitation of the existing
approaches is that they are at most
semi-automated~\cite{10.1145/3098954.3105821,9018111},
still requiring the analyst to assist
in their operation. Our goal in this work
is to present a fully-automated approach
for hidden path discovery and trace
analysis.

\noindent \textbf{Symbolic Execution at Scale.}
A major drawback of symbolic executors is that
they are very slow in comparison to traditional
sandboxes, which makes large-scale analysis hard.
The academic literature reports a case in which
the analysis of each one 60 thousand samples
considered in a study required an average of 
28 minutes~\cite{10.1007/978-3-319-75160-3_22}.
This often limits studies to few samples, such
as 50~\cite{Alsaleh2019}. Even when analysis
are scaled, such as in the Minesweeper's 
case~\cite{Brumley2008}, the provided analytic's
data fast become outdated as these studies are 
hardly ever repetead to cover newly developed 
logic bombs.

\noindent \textbf{\angr} is an open-source
powerful tool for binary
analysis~\cite{shoshitaishvili2016state}, thus
being selected as basis for the \tool development.
\angr was also used as basis for other research
work in multiple aspects, such as for: tracing
disjoint binary functions~\cite{9060117} 
(a technique presented in
~\cite{10.1145/1866307.1866354}), fixing
binary loading~\cite{8023121}, or in concolic
executions~\cite{gritti_symbion20}. Despite
powerful, \angr has some limitations, as
pointed in previous work~\cite{8552415, 10.1145/3243734.3243847}. Consequently,
these are also \tool's drawbacks.

\noindent \textbf{Other Symbolic Executors.} 
Besides \angr, other symbolic executors and
analyzers could provide similar support for 
\tool, such as \texttt{Metasm}~\cite{metasm},
\texttt{Miasm}~\cite{miasm}, and
\texttt{Triton}~\cite{triton}. A detailed
comparison of binary analysis frameworks is
presented in~\cite{10.1145/3359789.3359796}.

\section{Conclusion}
\label{sec:conc}

In this work, we investigated the problem of logic bomb detection and the inspection of 
context-sensitive malware. We observed a lack of automated solutions for these tasks and 
designed a system for assist analysts to tackle them. We proposed \tool, a solution able to 
inspect multiple execution paths via symbolic execution, whose goal is to discover triggers 
of malicious behaviors (in the form of function inputs and returns).
We designed and implemented \tool on top of \angr so it was able to
decompile functions that may be patched with the discovered symbolic
values. These patched functions then become ready to be injected in
the monitored process while the patched subject runs on analyses
sandboxes. We evaluated \tool using a set of Linux and Windows evasive
samples. \tool was able to produce automatic patches for the most
common evasion techiniques (e.g., \texttt{ptrace} and/or
\texttt{IsDebuggerPresent} checks).

\noindent \textbf{Reproducibility.} All code pieces
considered in this work are available in the repository:
\url{https://github.com/marcusbotacin/MalVerse}

\bibliographystyle{sbc}
\bibliography{references}

\end{document}